\documentclass[conference]{IEEEtran}
\IEEEoverridecommandlockouts
\usepackage{cite}
\usepackage{amsmath,amssymb,amsfonts}
\usepackage{graphicx}
\usepackage{textcomp}
\usepackage{xcolor}
\usepackage{algorithm}
\usepackage{algpseudocode}
\usepackage{multirow}
\usepackage{tikz} % To generate the plot from csv
\usepackage{pgfplots}
\usepackage{caption}
\usepackage{array}
\newcolumntype{P}[1]{>{\centering\arraybackslash}p{#1}}
\usepackage{url}
\usepackage{float}

\def\BibTeX{{\rm B\kern-.05em{\sc i\kern-.025em b}\kern-.08em
    T\kern-.1667em\lower.7ex\hbox{E}\kern-.125emX}}
\begin{document}

\title{Fully Neural Network Mode Based Intra Prediction of Variable Block Size\\
\thanks{This work was supported in part by JST, PRESTO Grant Number JPMJPR19M5, Japan.}
}

\author{\IEEEauthorblockN{ \textit{ Heming Sun$^\ast$$^\dagger$, Lu Yu$^\ddagger$, Jiro Katto$^\ast$$^\S$ } } \\
	\IEEEauthorblockA{{ $^\ast$Waseda Research Institute for Science and Engineering, Waseda University, Tokyo, Japan  } \\
		{ $^\dagger$JST, PRESTO, 4-1-8 Honcho, Kawaguchi, Saitama, Japan  } \\
		{ $^\ddagger$Institute of Information and Communication Engineering, Zhejiang University, Hangzhou, China  } \\
		{ $^\S$Department of Computer Science and Communication Engineering, Waseda University, Tokyo, Japan  }}}

\IEEEoverridecommandlockouts
\IEEEpubid{\makebox[\columnwidth]{978-1-7281-8068-7/20/\$31.00~\copyright2020 IEEE \hfill} \hspace{\columnsep}\makebox[\columnwidth]{ }}

\maketitle

\IEEEpubidadjcol

\begin{abstract}
Intra prediction is an essential component in the image coding. This paper gives an intra prediction framework completely based on neural network modes (NM). Each NM can be regarded as a regression from the neighboring reference blocks to the current coding block. (1) For variable block size, we utilize different network structures. For small blocks 4$\times$4 and 8$\times$8, fully connected networks are used, while for large blocks 16$\times$16 and 32$\times$32, convolutional neural networks are exploited. (2) For each prediction mode, we develop a specific pre-trained network to boost the regression accuracy. When integrating into HEVC test model, we can save 3.55\%, 3.03\% and 3.27\% BD-rate for Y, U, V components compared with the anchor. As far as we know, this is the first work to explore a fully NM based framework for intra prediction, and we reach a better coding gain with a lower complexity compared with the previous work.

\end{abstract}


\section{Introduction}
Intra prediction is used to remove the spatial redundancy in the image/video coding. For each block, the predicted pixels are the linearly interpolated results based on the neighboring pixels. To enhance the prediction accuracy, more prediction modes and block sizes have been developed in the past standards. For the modes, the amount has been extended from 9 in H.264 \cite{wiegand2003overview} to 35 in HEVC \cite{sullivan2012overview}. For the block size, the largest block has been enlarged from 16$\times$16 in H.264 to 64$\times$64 in HEVC. Though providing more prediction modes and blocks can improve the coding gain, there is a limitation especially for the blocks without explicit directional texture.

So far, all the standards utilized a single reference line composed of upper and left pixels. To further extract the correlation between adjacent pixels, multiple reference lines are used. As reported in \cite{li2016efficient}, using more reference pixels could reach up to 4.3\% coding gain. In addition, to obtain the non-linearity and a high-level relationship between reference blocks and current block, deep learning has been used for intra prediction in several literature. Li \textit{et al.} \cite{li2018fully} exploited fully connected (FC) networks for various block sizes from 4$\times$4 to 32$\times$32. Dumas \textit{et al.} \cite{dumas2019context} stated that using convolutional neural network (CNN) performs better than FC for blocks larger than 8$\times$8. Hu \textit{et al.} \cite{hu2019progressive} presented a new structure based on recurrent neural network (RNN). Sun \textit{et al.} \cite{sun2020enhanced} studied different combination schemes of traditional modes (TM) and neural network modes (NM) for the fixed block 8$\times$8. Zhu \textit{et al.} \cite{zhu2019generative} regarded intra prediction as an inpainting problem and provided 35 NMs for 64$\times$64.

Though the previous works have achieved significant coding gain, there are still some remaining problems. First is that TM still remains in the coding framework. In \cite{li2018fully}, \cite{dumas2019context} and \cite{hu2019progressive}, one or two NMs were provided. In \cite{sun2020enhanced}, at most seven NMs were exploited. As a result, the hybrid mode handling of TM and NM might lead to biased selection to one of them. Second is that the complexity is extremely high for some networks such as \cite{zhu2019generative} which is more than 5000x decoding complexity.

In this paper, different from the hybrid mode handling in previous works, we explore a fully NM based intra coding framework. The contributions are listed in the follows.

\begin{itemize}
	
\item For all the 35 modes of variable block sizes from 4$\times$4 to 32$\times$32, we propose a corresponding network model.
	
\item For the small blocks 4$\times$4 and 8$\times$8, FC networks are presented. For the large blocks 16$\times$16 and 32$\times$32, CNN are used.
	
\item We select the optimized model by exploring the coding gain and complexity, and analyze the probability of the best NM for the mode signaling.

\end{itemize}

\section{Traditional and Learned Intra Prediction}

\subsection{HEVC Intra Prediction}

HEVC provides 35 TMs, which can be categorized to non-directional and directional. Non-directional TMs are composed of Planar and DC. For the 33 directional TMs, the prediction is performed according to the predictive angles. About the mode signaling of 35 TMs, first, one \textit{bin} is consumed to represent whether the mode is among the most probable mode (MPM) set or not. If so, one or two \textit{bins} are used to indicate the MPM index. Otherwise, five \textit{bins} are cost to represent the mode among the remaining 32 TMs as shown in Table \ref{table_mode_signaling}.

\subsection{Learned Intra Prediction with Appending and Substitution Scheme}

When introducing NMs, there are two schemes to integrate NMs with TMs as described in \cite{sun2020enhanced}. One is appending scheme that is to append NMs to all the 35 TMs. In this case, additional flags are required to signal the new modes. The other is substituting TMs by NMs. In this case, since the overall number of modes is the same as origin, the signaling scheme follows Table \ref{table_mode_signaling}. As reported in \cite{dumas2019context} and \cite{sun2020enhanced}, the coding gain by appending scheme is obviously larger than the substitution scheme. However, the substitution scheme is only limited to at most three modes in \cite{sun2020enhanced}. In this work, we extend the number from 3 to 35 to create a fully NM-based framework.

\section{Proposed Fully Neural Network Mode Based Intra Coding Framework}

\subsection{Network Structure Analysis}\label{AA}
\begin{figure}[t]
	\centerline{\includegraphics[height=1.8cm]{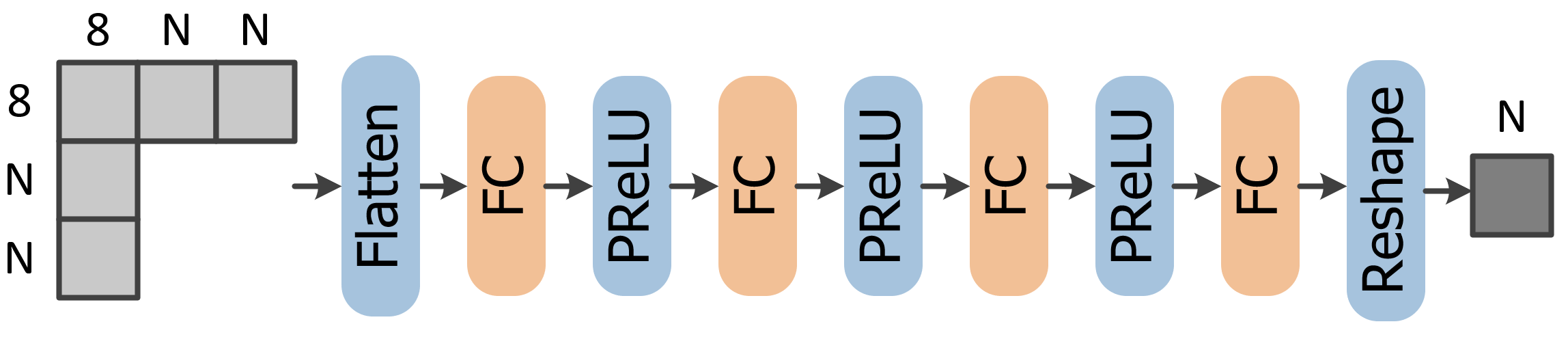}}
	
	\captionsetup{font={small}}
	\caption{Network for block 4$\times$4 and 8$\times$8. N is the block size.}
	\label{fig_fc_network}
	\vspace{-5mm}
\end{figure}

As described in \cite{dumas2019context}, FC and CNN are suitable for smaller and larger blocks respectively. Motivated by \cite{dumas2019context}, we also use FC network for  4$\times$4 and 8$\times$8 as shown in Fig. \ref{fig_fc_network}. First, the neighboring references blocks are flattened to one-dimension vector with (4$\times$N+8)$\times$8 nodes. By passing through four FC layers, we reshape the one-dimension vector to two-dimension N$\times$N block. The number of nodes for each layer is determined according to the analysis of the coding gain and complexity. The coding gain is evaluated by PSNR and the complexity is measured by FLOPs. The results are shown in Table \ref{table_network}. We first train a baseline heavy model with 512 nodes, and then reduce the number of nodes by half. When reducing the number of nodes to 256 and 128, the coding loss is small. However, when further reducing the dimension to 64, there is an obvious coding loss that is 0.21dB (25.03dB-24.82dB) for 4$\times$4 and 0.34dB (27.08dB-26.74dB) for 8$\times$8. Therefore, we select the node as 128.

\begin{table}[t]
	\setlength{\abovecaptionskip}{-3pt}
	\captionsetup{font={small}}
	\caption{Mode signaling for 35 TMs}
	\begin{center}
		
		\resizebox{0.90\width}{!}{
		
		\begin{tabular}{c|c|c|c|c}
			\hline
			\textbf{Mode} 			& \textbf{MPM$_0$}	& \textbf{MPM$_1$} 	& \textbf{MPM$_2$} 	 	& \textbf{Non-MPM} 	\\ 			\hline	
			\textbf{Codeword} 		& 10				& 110 				& 111		 			& 0\{5\textit{bins}\}		\\ 			\hline	
		\end{tabular}
	
		}
	
		\label{table_mode_signaling}
	\end{center}
\vspace{-2mm}
\end{table}

\begin{table}[t]
	
	\setlength{\abovecaptionskip}{-3pt}
	\captionsetup{font={small}}
	\caption{Select the number of node/filter based on the trade-off between coding gain (PSNR) and complexity (FLOPs)}
	\begin{center}
		
		\resizebox{0.90\width}{!}{
		
		\begin{tabular}{c|c|c|c|c|c}
			\hline
			\textbf{ }   					& \text{Node/Filter}		& 512 	& 256 	& \textbf{128} 	& 64   			\\ 
			\cline{2-6}	
			$ \textbf{TU 4$\times$4}   $ 	& \text{PSNR (dB)}			& 25.09	& 25.06	& \textbf{25.03}	& 24.82			\\ 			
			\cline{2-6}	
			\textbf{ } 						& \text{FLOPs (\textbf{K})}			& 1270 	& 373 	& \textbf{121} 	& 44			\\ 			
			\hline	
			\hline										
			\textbf{ } 						& \text{Node/Filter}		& 512 	& 256 	& \textbf{128} 	& 64   			\\ 		
			\cline{2-6}
			$ \textbf{TU 8$\times$8} 	 $	& \text{PSNR (dB)}			& 27.26	& 27.20	& \textbf{27.08}	& 26.74 		\\ 			
			\cline{2-6}
			\textbf{ }						& \text{FLOPs (\textbf{K})}			& 1450	& 464	& \textbf{167}	& 67   			\\ 			
			\hline		
			\hline														
			\textbf{ }						& \text{Node/Filter}		& 64  	& 32 	& \textbf{16} 	& 8 			\\			
			\cline{2-6}
			$ \textbf{TU 16$\times$16}   $	& \text{PSNR (dB)}			& 28.24	& 28.25	& \textbf{28.13}	& 27.93 		\\ 			
			\cline{2-6}
			\textbf{ }						& \text{FLOPs (\textbf{M})}			& 97.5 	& 24.5 	& \textbf{6.4}	& 1.7  			\\ 			
			\hline			
			\hline													
			\textbf{ } 						& \text{Node/Filter}		& 64  	& 32 	& \textbf{16} 	& 8 			\\ 			
			\cline{2-6}
			$ \textbf{TU 32$\times$32}   $	& \text{PSNR (dB)}			& 29.44 & 29.37	& \textbf{29.41}	& 29.29 		\\ 			
			\cline{2-6}
			\textbf{ } 						& \text{FLOPs (\textbf{M})}			& 547.2	& 138.4	& \textbf{35.4}	& 9.2 	 	 	\\ 			
			\hline
		\end{tabular}
	
		}
	
		\label{table_network}
	\end{center}
\vspace{-5mm}
\end{table}

\begin{figure}[t]
	\centerline{\includegraphics[height=2.7cm]{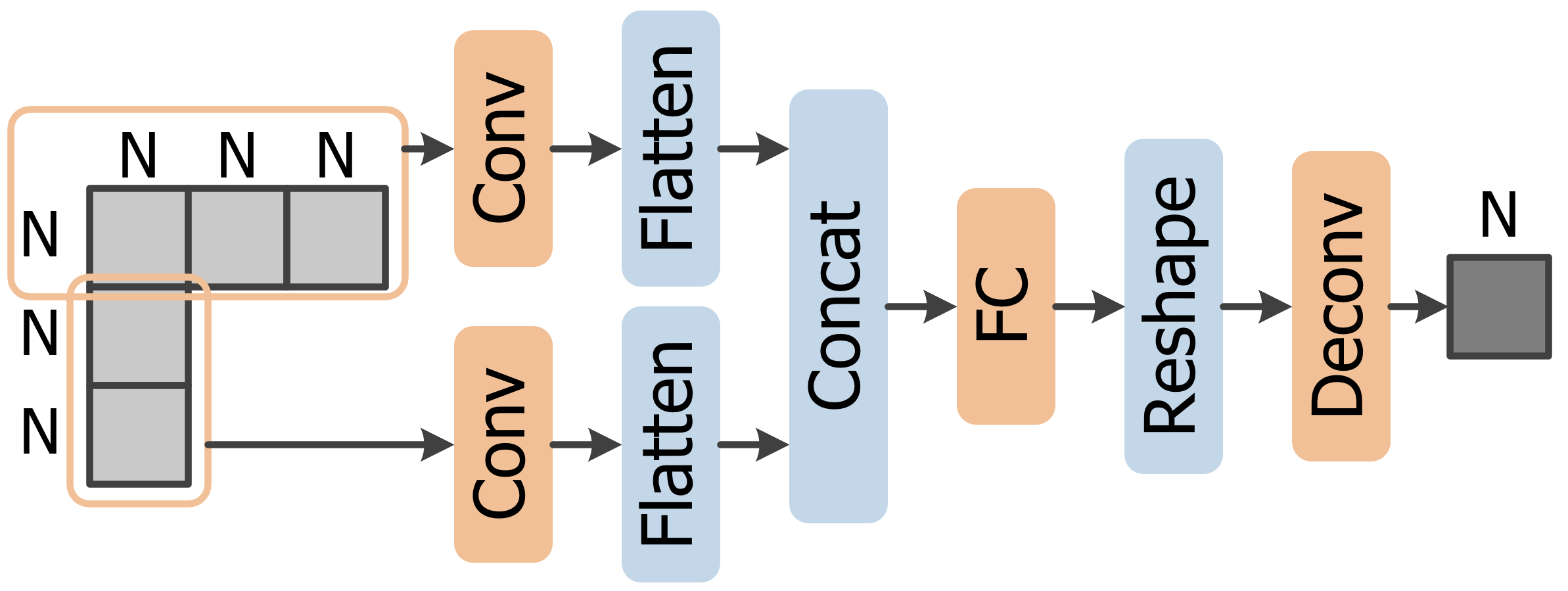}}
	
	\captionsetup{font={small}}
	\caption{Network for block 16$\times$16 and 32$\times$32. N is the block size.}
	\label{fig_cnn_network}
	\vspace{-5mm}
\end{figure}

\begin{table}[t]
	
	\setlength{\abovecaptionskip}{-3pt}
	\captionsetup{font={small}}
	\caption{Convolutional layer structures for 16$\times$16 and 32$\times$32}
	\begin{center}
		
		\resizebox{0.90\width}{!}{
		
		\begin{tabular}{P{1.00cm}|P{0.70cm}|P{1.00cm}|P{0.70cm}|P{0.70cm}|P{1.00cm}|P{0.70cm}}
			\hline
			\text{} & \multicolumn{3}{c|}{\textbf{16$\times$16}} & \multicolumn{3}{c}{\textbf{32$\times$32}} \\
			\hline			
			\textbf{Layer number}  	& \textbf{Kernel size}		& \textbf{Filter number}	& \textbf{Stride} 		& \textbf{Kernel size}		& \textbf{Filter number}	& \textbf{Stride} 	\\ 			
			\hline
			\hline
			1 				& 5$\times$5	&F	& 2 	& 5$\times$5	&F	& 2 		\\ 			
			\hline
			2				& 3$\times$3	&F	& 1  	& 5$\times$5	&2F	& 2 		\\ 			
			\hline
			3 				& 5$\times$5	&2F	& 2  	& 5$\times$5	&4F	& 2 			\\ 			
			\hline
			4				& 3$\times$3	&2F	& 1  	& 5$\times$5	&8F	& 2  		\\ 			
			\hline
			5				& -				& -	& -  	& 3$\times$3	&8F	& 1   		\\ 			
			\hline			
		\end{tabular}
	
		}
	
		\label{table_conv}
	\end{center}
\vspace{-5mm}
\end{table}

For larger blocks, we utilize a CNN network as shown in Fig. \ref{fig_cnn_network}. To keep the spatial information, the above three blocks and the left two blocks are sent to two separate convolutional paths. The convolutional path is composed of several convolutional layers as shown in Table \ref{table_conv}. For each path, we conduct the down-sampling to obtain the latent information, and then flatten to one-dimensional vector. Two vectors are concatenated and then pass a FC layer. The number of outputs nodes of the FC layer is 1/5 of the input nodes. Finally, we reshape to two-dimension and use deconvolutional layers to up-sample to the original block size N$\times$N. The selections of filters and strides are shown in Table \ref{table_conv}. We use four and five convolutional layers for 16$\times$16 and 32$\times$32, respectively. Different from \cite{dumas2019context}, we utilize PReLU as the activation function. The number of filters $F$ is selected as 16 for 16$\times$16 and 32$\times$32 as a trade-off between coding gain and complexity as shown in Table \ref{table_network}.

\begin{figure}[t]
%Ex6:plot from data

\centering

\resizebox{0.75\width}{!}{

\begin{tikzpicture}
\begin{axis}[
label style={font=\small},
tick label style={font=\small} ,
xlabel={Mode},
ylabel={ PSNR (dB) },
xmin=0, xmax=35,
ymin=23, ymax=34,
xtick={0,5,10,15,20,25,30,35},
minor y tick num = 1,
legend pos=south west,
ymajorgrids=true,
grid style=dashed,
legend style={font=\fontsize{7}{7}\selectfont, row sep=-3pt},
legend entries = { TM, NM (FC\_1024node), NM (CNN\_16filter) },
]

\addplot[ color=green, mark=*, mark size=1.5pt] table {hevc_tm.dat};
\addplot[ color=blue, mark=*, mark size=1.5pt] table {16_fc_4x1024.dat};
\addplot[ color=red, mark=*, mark size=1.5pt] table {16_inria_4x16.dat};

\end{axis}

\end{tikzpicture}

}

\captionsetup{font={small}}
\caption{PSNR comparison of FC and CNN network for 35 modes of 16$\times$16.}
\label{fig_psnr_compare}

\vspace{-5mm}
\end{figure}
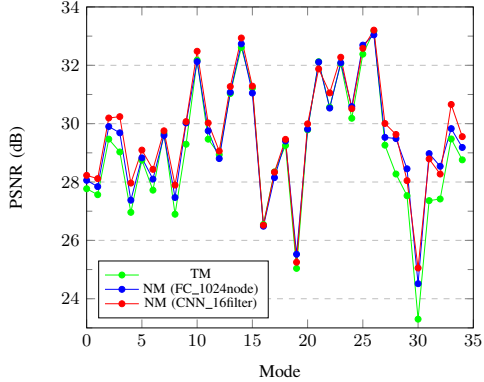

Though \cite{dumas2019context} has concluded that using convolutional layers could help the prediction of larger block such as 16$\times$16, the conclusion was drawn when utilizing one NM. To ensure that this conclusion also works for all the 35 NMs, we adopt a heavy FC network with 1024 nodes, and the PSNR comparison between FC and CNN network for 16$\times$16 is shown in Fig. \ref{fig_psnr_compare}. We can see that even using a powerful FC model with large nodes, the average performance is still worse than CNN model in most scenarios. In addition, CNN model is better than the original TM in almost all the cases.

\subsection{Coding Framework with Fully Neural Network Modes}

There are overall 35 NMs, we select the best NM by the following steps. First, several candidate modes are selected by sum of absolute transformed differences (SATD) cost. Eight candidates are picked up for block 4$\times$4 and 8$\times$8, while three candidates are chosen for the other blocks. In addition to the candidate modes selected by SATD, MPMs are also appended in the candidate mode list. After creating the candidate mode list, the best NM is determined by comparing the R-D cost.

The mode signaling for the 35 TM in Table \ref{table_mode_signaling} was designed based on probability analysis. Modes with higher probability to be the best mode will be allocated fewer \textit{bins}. Given that we have two following equations
\begin{equation}
\small
P( M=BM | M \in MPM) > P( M=BM | M \in Non-MPM))
\label{eq_1}
\end{equation}
\begin{equation}
\small
P( MPM_0=BM ) > P( M=BM | M \in \{MPM_1, MPM_2\}))
\label{eq_2}
\end{equation}

\noindent where BM is the best mode, thus MPMs are allocated fewer \textit{bins} compared with Non-MPMs. Among MPMs, MPM$_0$ is allocated one fewer \textit{bin} than the other two MPMs.

However, Eq. \ref{eq_1} and Eq. \ref{eq_2} are for TM. To ensure that this mode signaling is applicable to 35 NM, we analyze the probability for the sequence \textit{RaceHorses} at QP32, and the results are shown in Table \ref{table_mpm_ratio}. We can see that MPM still owns larger probability than non-MPM in our proposal, which is 59.9\% against 40.1\%. Among MPMs, MPM$_0$ has 29.2\% to be the best mode which is higher than MPM$_1$ and MPM$_2$. We also evaluate the probability of TMs being the best mode in the original case. We can see that there is no obvious bias between the probability distribution of our proposal and origin.

\begin{figure}[t]
	%Ex6:plot from data
	
	\centering
	
	\resizebox{0.75\width}{!}{
	\begin{tikzpicture}
	\begin{axis}[
	label style={font=\small},
	tick label style={font=\small} ,
	ymode=log,
	x label style={at={(axis description cs:0.5,0.0)}},
	y label style={at={(axis description cs:0.05,.5)},anchor=south},
	xlabel={Mode},
	ylabel={ Probability (log scale) },
	xmin=0, xmax=35,
	ymin=0, ymax=40,
	xtick={0,1,5,10,15,20,26,30,35},
	minor y tick num = 1,
	legend pos=north east,
	ymajorgrids=true,
	grid style=dashed,
	legend style={font=\fontsize{8}{8}\selectfont, row sep=-3pt},
	legend entries = { Origin (35 TMs), Proposal (35 NMs)  },
	]
	\addplot[ color=blue, mark=*, mark size=1.5pt] table {best_tm.dat};
	\addplot[ color=red, mark=*, mark size=1.5pt] table {best_nm.dat};
	
	\end{axis}
	\end{tikzpicture}
	
	}
	\captionsetup{font={small}}
	\caption{Probability of 35 NMs and TMs to become the best mode.}
	\label{fig_best_mode}
\vspace{-4mm}
\end{figure}
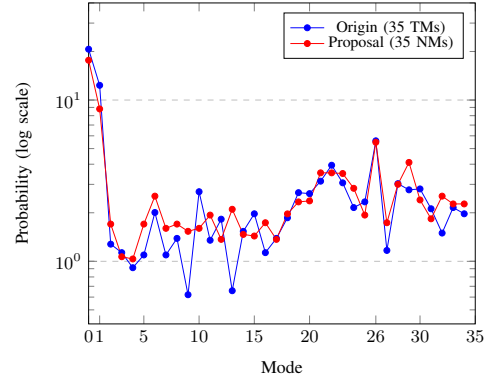

About the composition of the MPM, when MPM$_0$ and MPM$_1$ are not same, we set MPM$_2$ as one of Planar (mode 0), DC (mode 1) and Vertical (mode 26) considering that these three modes have the highest probability to be the best mode. Therefore, we also evaluate the probability of all the 35 NMs to be the best mode in Fig. \ref{fig_best_mode}. We can see that mode 0, 1 and 26 still own the highest probability in our proposals.

\begin{table}[t]
	\setlength{\abovecaptionskip}{-3pt}
	\captionsetup{font={small}}
	\caption{Probability to be the best mode for MPMs and Non-MPMs in the case of using TM and NM}
	\begin{center}
		
		\resizebox{0.90\width}{!}{
		
		\begin{tabular}{c|c|c|c|c}
			\hline
			\textbf{ }		& \textbf{MPM$_0$} 		& \textbf{MPM$_1$}		& \textbf{MPM$_2$} 		& \textbf{ $\bigcup$ \{32 Non-MPMs\} } 	   	\\ 			\hline
			\textbf{Origin (35 TMs)}				&29.5 					& 18.7					& 14.9 				& 36.9 					\\ 			\hline
			\textbf{Proposal (35 NMs)}				&29.2 					& 16.7					& 14.0 				& 40.1 					\\ 			\hline
		\end{tabular}
	
		}
		\vspace{-3mm}
		\label{table_mpm_ratio}
	\end{center}
\end{table}

\section{Training Process}\label{s4}

The training is a regression problem from reference blocks $R$ to original block $Y$ with the network parameter $\theta$.  The loss function is composed of MSE and a regularization term as given in Eq. \ref{eq_loss}
\begin{equation}
\emph{J}(\theta) = \dfrac{1}{M} \sum_{m=0}^{M-1} ||F(R^m, \theta)-Y^m||_2 + \lambda||\theta_w||_2^2
\label{eq_loss}
\end{equation}
\noindent where $\lambda$ is set as 0.0005 and the batch size $M$ is 16. We used the New York city library \cite{wilson2014robust} as the training set. We encode each image with four QPs (22, 27, 32, 37), and the best block is used for the training set $\Gamma = \bigcup_0^{34} \Gamma_i $ where $i$ is the mode number and $\Gamma_i$ is composed of the blocks which select $i$ as the best mode. First, we train one baseline model based on all the training set $\Gamma$.  After that, we fine tune the baseline model for each specific mode $i$ from 0 to 34 based on the corresponding training set $\Gamma_i$. The training procedures are shown in \textit{Algorithm 1}. ADAM \cite{kingma2014adam} is used as the optimizer and the learning rate is set as 0.0001 and 0.0004 for FC and CNN network respectively.

\begin{algorithm} [H]
	\small
	\caption{Proposed Training Method for 35 NMs}
	\label{alg:Framwork}
	\begin{algorithmic}[1]
		
		\For{ training iterations \textit{$I_1$} = $\tfrac{card( \Gamma )}{M}$ }
		
		\State Update $\theta$ with ADAM optimizer
		
		\EndFor
		
		\State Obtain baseline model with $\theta$
		
		\For { each mode $i$ }
		
		\For { fine tuning iterations \textit{$I_2$} = $\tfrac{card( \Gamma_i )}{M}$ }
		
		\State Update $\theta_i$ with ADAM optimizer
		
		\EndFor
		
		\State \textbf{return} $\theta_i$
		
		\EndFor
		
	\end{algorithmic}
\end{algorithm}

\section{Experimental Results}

\subsection{Coding Gain Analysis}
We integrate the proposal into HM 16.9 \cite{hm16}, and test the first frame of the sequences in Table \ref{table_bd-rate_sequence}. We strictly follow the coding configuration of "all-intra main" given by HEVC common test condition (CTC) \cite{ctc}. Four QPs (22, 27, 32, 37) are tested, and the coding efficiency comparison with the anchor (original HM16.9) is measured by BD-rate \cite{bjontegaard2001calculation}. It should be noted that there is no overlap between training set \cite{wilson2014robust} and test set \cite{ctc}.

The results of BD-rate are shown in Table \ref{table_bd-rate_sequence}. The results show that we can save BD-rates for all the test sequences. On average, 3.55\%, 3.03\% and 3.27\% Y, U, V BD-rate can be saved compared with the anchor. Compared with \cite{li2018fully}, we can achieve large BD-rate reduction for all the three channels. We also compare the class-level results to the three previous works with the same-level complexity in Table \ref{table_bd-rate_class}. When using the proposed model, we can also achieve best coding gain at Class B and E among all the works.

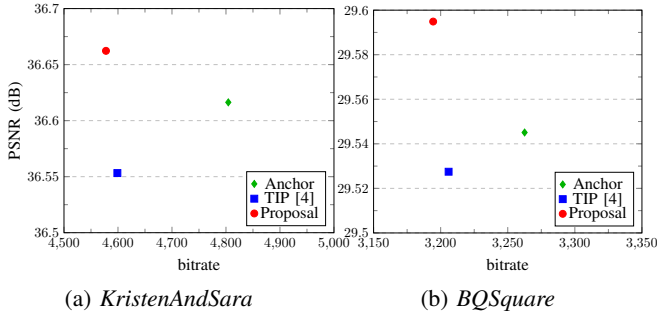
\begin{figure}[t]
	%Ex6:plot from data
	
	\begin{minipage}[b]{0.48\linewidth}
	
	\resizebox{!}{0.4\width}
	{
		
		\begin{tikzpicture}
		\begin{axis}
		%\addplot+[only marks]
		[
		label style={font=\large},
		tick label style={font=\normalsize} ,
		xlabel={bitrate},
		ylabel={ PSNR (dB) },
		xmin=4500, xmax=5000,
		ymin=36.5, ymax=36.7,
		minor x tick num = 1,
		minor y tick num = 1,
		legend pos=south east,
		ymajorgrids=true,
		grid style=dashed,
		legend style={font=\fontsize{12}{12}\selectfont, row sep=-3pt},
		legend entries = { Anchor, TIP\cite{li2018fully}, Proposal },
		]
		\addplot[ only marks, color=green!70!black, mark=diamond*, mark size=2.5pt] table {kas_orig_rd_qp37.dat};
		\addplot[ only marks, color=blue, mark=square*, mark size=2.5pt] table {kas_tip_rd_qp37.dat};
		\addplot[ only marks, color=red, mark=*, mark size=2.5pt] table {kas_proposal_rd_qp37.dat};
		
		\end{axis}
		\end{tikzpicture}
	}
	\centerline{\small (a) \textit{KristenAndSara}}
	\end{minipage}
	\begin{minipage}[b]{0.48\linewidth}
	
	\resizebox{!}{0.42\width}
	{
		
		\begin{tikzpicture}
		\begin{axis}
		%\addplot+[only marks]
		[
		label style={font=\large},
		tick label style={font=\normalsize} ,
		xlabel={bitrate},
		xmin=3150, xmax=3350,
		ymin=29.5, ymax=29.6,
		minor x tick num = 1,
		minor y tick num = 1,
		legend pos=south east,
		ymajorgrids=true,
		grid style=dashed,
		legend style={font=\fontsize{12}{12}\selectfont, row sep=-3pt},
		legend entries = { Anchor, TIP\cite{li2018fully}, Proposal },
		]		
		\addplot[ only marks, color=green!70!black, mark=diamond*, mark size=2.5pt] table {bqsquare_orig_rd_qp37.dat};
		\addplot[ only marks, color=blue, mark=square*, mark size=2.5pt] table {bqsquare_tip_rd_qp37.dat};
		\addplot[ only marks, color=red, mark=*, mark size=2.5pt] table {bqsquare_proposal_rd_qp37.dat};
		
		\end{axis}
		\end{tikzpicture}
	}
	\centerline{\small (b) \textit{BQSquare}}
	\end{minipage}
	\captionsetup{font={small}}
	\caption{ R-D comparison at low bitrates (QP37). \cite{li2018fully} achieved smaller bitrates than the anchor at the cost of worse PSNR, while we can achieve both smaller bitrates and better PSNR. }
	\label{fig_rd_curve}
\vspace{-3mm}
\end{figure}

To clarify the relationship of bitrate and PSNR when using the proposal, we give the R-D results of QP37 of the sequence \textit{KristenAndSara} and \textit{BQSquare} in Fig. \ref{fig_rd_curve}. We can see that we can save bitrates when achieving better PSNR compared with the anchor. The method \cite{li2018fully} can also reduce the bitrate. However, there is some PSNR loss. It is because \cite{li2018fully} appended NMs. In the case of NM being the best mode, the number of bit consumption for the mode signaling can be significantly reduced while the coding quality is also degraded.

\begin{table}[t]
	
	\setlength{\abovecaptionskip}{-3pt}   %调整图片标题与图距离	
	\captionsetup{font={small}}
	\caption{Coding gain comparison with previous work}
	\scriptsize
	\begin{center}
		\begin{tabular}{P{0.40cm}|c|P{0.4cm}|P{0.4cm}|P{0.4cm}|P{0.55cm}|P{0.55cm}|P{0.55cm}}
			\hline
			\multirow{2}*{\textbf{Class}} & \multirow{2}*{\textbf{Sequence}}				&\multicolumn{3}{c|}{\textbf{TIP \cite{li2018fully}}}		&\multicolumn{3}{c}{\textbf{Proposal}} \\
			\cline{3-8}
			\text{ } & \text{ }		& \textbf{Y} &\textbf{U}&\textbf{V} 	& \textbf{Y} &\textbf{U}&\textbf{V} \\
			\hline
			\hline
			\multirow{8}*{\textbf{A}}		&\text{Tango}		&\text{-7.4} 	&\text{-0.8}	&\text{-4.3}	&\text{-7.08} 	&\text{-7.27}	&\text{-6.74}\\
			\cline{2-8} 
			\text{ }		&\text{Drums100}					&\text{-3.8} 	&\text{-1.6}	&\text{-1.6}	&\text{-2.94} 	&\text{-3.25}	&\text{-3.15}\\
			\cline{2-8} 
			\text{ }		&\text{CampfireParty}				&\text{-3.0} 	&\text{-3.3}	&\text{-3.0}	&\text{-1.37} 	&\text{-1.99}	&\text{-1.51}\\
			\cline{2-8} 
			\textbf{}	&\text{ToddlerFountain}					&\text{-3.4} 	&\text{2.8}		&\text{-1.5}	&\text{-2.75} 	&\text{0.14}	&\text{-1.71}\\
			\cline{2-8} 
			\text{ }		&\text{CatRobot}					&\text{-4.2} 	&\text{-2.4}	&\text{-2.6}	&\text{-4.13} 	&\text{-3.23}	&\text{-3.58}\\
			\cline{2-8} 
			\text{ }		&\text{TrafficFlow}					&\text{-4.2} 	&\text{-1.3}	&\text{-1.3}	&\text{-4.62} 	&\text{-2.89}	&\text{-2.41}\\
			\cline{2-8} 
			\text{ }		&\text{DaylightRoad}				&\text{-4.5} 	&\text{0.1}		&\text{-1.8}	&\text{-4.85} 	&\text{-3.49}	&\text{-4.15}\\
			\cline{2-8} 
			\textbf{ }		&\text{Rollercoaster}				&\text{-5.8} 	&\text{-3.7}	&\text{-2.7}	&\text{-5.11} 	&\text{-4.48}	&\text{-4.45}\\
			\hline
			\multicolumn{2}{c|}{\textbf{Average of Class A}}	&\textbf{-4.5} 	&\textbf{-1.3}	&\textbf{-2.4}	&\textbf{-4.11} 	&\textbf{-3.31}		&\textbf{-3.46}\\
			\hline
			\hline					
			\multirow{5}*{\textbf{B}}		&\text{Kimono}		&\text{-3.1} 	&\text{-2.1}	&\text{-1.5}	&\text{-2.58} 	&\text{-3.25}	&\text{-2.91}\\
			\cline{2-8} 
			\text{ }	&\text{ParkScene}						&\text{-3.6} 	&\text{-2.2}	&\text{-2.4}	&\text{-2.58} 	&\text{-2.33}	&\text{-2.23}\\
			\cline{2-8} 
			\text{ }		&\text{Cactus}						&\text{-3.2} 	&\text{-1.8}	&\text{-1.5}	&\text{-4.27} 	&\text{-1.82}	&\text{-4.10}\\
			\cline{2-8} 
			\textbf{ }		&\text{BQTerrace}					&\text{-2.1} 	&\text{-1.3}	&\text{-0.5}	&\text{-4.72} 	&\text{-2.41}	&\text{-2.59}\\
			\cline{2-8} 
			\textbf{ }		&\text{BasketballDrive}				&\text{-3.6} 	&\text{-2.9}	&\text{-2.7}	&\text{-2.48} 	&\text{-1.53}	&\text{-2.77}\\
			\hline
			\multicolumn{2}{c|}{\textbf{Average of Class B}}	&\textbf{-3.1} 	&\textbf{-2.1}	&\textbf{-1.7}	&\textbf{-3.33} 	&\textbf{-2.27}	&\textbf{-2.92}\\
			\hline
			\hline								
			\multirow{4}*{\textbf{C}}	&\text{BasketballDrill}	&\text{-1.5} 	&\text{-3.3}	&\text{-2.2}		&\text{-2.70} 	&\text{-4.75}	&\text{-4.28}\\
			\cline{2-8} 
			\text{ }	&\text{BQMall}							&\text{-2.2} 	&\text{-1.9}	&\text{-1.0}		&\text{-2.03} 	&\text{-2.53}	&\text{-3.14}\\
			\cline{2-8} 
			\text{ }		&\text{PartyScene}					&\text{-1.6} 	&\text{-1.2}	&\text{-0.1}		&\text{-1.81} 	&\text{-1.62}	&\text{-1.20}\\
			\cline{2-8} 
			\textbf{ }		&\text{RaceHorsesC}					&\text{-3.2} 	&\text{-1.9}	&\text{-2.8}		&\text{-3.01} 	&\text{-1.98}	&\text{-2.80}\\
			\hline
			\multicolumn{2}{c|}{\textbf{Average of Class C}}	&\textbf{-2.1} 	&\textbf{-2.1}	&\textbf{-1.5}		&\textbf{-2.39} 	&\textbf{-2.72}		&\textbf{-2.85}\\
			\hline
			\hline								
			\multirow{4}*{\textbf{D}}	&\text{BasketballPass}	&\text{-1.2} 	&\text{-0.3}	&\text{1.1}			&\text{-2.32} 	&\text{-2.01}	&\text{-1.31}\\
			\cline{2-8} 
			\text{}	&\text{BQSquare}							&\text{-0.9} 	&\text{-0.1}	&\text{-2.8}		&\text{-1.77} 	&\text{0.24}	&\text{-2.34}\\
			\cline{2-8} 
			\text{}		&\text{BlowingBubbles}					&\text{-1.9} 	&\text{-2.8}	&\text{-3.5}		&\text{-1.76} 	&\text{-4.23}	&\text{-3.36}\\
			\cline{2-8} 
			\textbf{ }		&\text{RaceHorses}					&\text{-3.2} 	&\text{-2.6}	&\text{-2.8}		&\text{-3.52} 	&\text{-4.08}	&\text{-4.38}\\
			\hline
			\multicolumn{2}{c|}{\textbf{Average of Class D}}	&\textbf{-1.8} 	&\textbf{-1.5}	&\textbf{-2.0}		&\textbf{-2.34  } 	&\textbf{-2.52}		&\textbf{-2.85}\\
			\hline
			\hline								
			\multirow{3}*{\textbf{E}}	&\text{FourPeople}		&\text{-4.4} 	&\text{-4.5}	&\text{-3.1}		&\text{-5.81} 	&\text{-5.05}		&\text{-4.81}\\
			\cline{2-8} 
			\text{}		&\text{Johnny}							&\text{-5.3} 	&\text{-3.1}	&\text{-3.5}		&\text{-5.33} 	&\text{-4.06}		&\text{-3.75}\\
			\cline{2-8} 
			\textbf{ }		&\text{KristenAndSara}				&\text{-3.9} 	&\text{-3.0}	&\text{-3.1}		&\text{-5.55} 	&\text{-4.91}		&\text{-4.70}\\
			\hline
			\multicolumn{2}{c|}{\textbf{Average of Class E}}	&\textbf{-4.5} 	&\textbf{-3.5}	&\textbf{-3.2}		&\textbf{-5.56} 	&\textbf{-4.67}		&\textbf{-4.42}\\
			\hline
			\hline								
			\multicolumn{2}{c|}{\textbf{Average of All Squences}}	&\textbf{-3.4} 	&\textbf{-1.9}		&\textbf{-2.1}		&\textbf{-3.55} 	&\textbf{-3.03}		&\textbf{-3.27}\\
			\hline
		\end{tabular}
	\end{center}
	\label{table_bd-rate_sequence}
	\vspace{-2mm}
	
\end{table}

\begin{table}[t]
	
	\setlength{\abovecaptionskip}{-3pt}   %调整图片标题与图距离	
	\captionsetup{font={small}}
	\caption{Class-level Y-BD-rate comparison with previous works}
	\begin{center}
		
		\resizebox{0.90\width}{!}{
		
		\begin{tabular}{P{1.00cm}|c|c|c|c}
			\hline			
			\textbf{ }				&\textbf{TIP \cite{li2018fully}}	& \textbf{TIP \cite{dumas2019context}}	& \textbf{TMM \cite{hu2019progressive}}	& \textbf{Proposed}	\\
			\hline
			\hline
			\textbf{Class A}		&\textbf{-4.5}		& \text{N/A}		& \text{N/A}			& \text{-4.11}  \\
			\hline
			\textbf{Class B}		&\text{-3.1}		& \text{-3.24}		& \text{-2.39}			& \textbf{-3.33}  \\
			\hline
			\textbf{Class C}		&\text{-2.1}		& \textbf{-3.09}		& \text{-2.31}			& \text{-2.39}  \\
			\hline
			\textbf{Class D}		&\text{-1.8}		& \textbf{-2.81}		& \text{-2.54}			& \text{-2.34}  \\
			\hline
			\textbf{Class E}		&\text{-4.5}		& \text{N/A}		& \text{-3.68}			& \textbf{-5.56}  \\
			\hline	
		\end{tabular}
	
		}
	
	\end{center}
	\label{table_bd-rate_class}
	\vspace{-1mm}
	
\end{table}

\begin{table}[t]
	
	\setlength{\abovecaptionskip}{-3pt}   %调整图片标题与图距离	
	\captionsetup{font={small}}
	\caption{Coding complexity comparison with previous works}
	\begin{center}
		
		\resizebox{0.90\width}{!}{
		
		\begin{tabular}{P{0.50cm}|P{1.60cm}|P{1.60cm}|P{1.60cm}|P{1.00cm}}
			\hline			
			\textbf{ }				&\textbf{TIP\cite{li2018fully}}	& \textbf{TIP\cite{dumas2019context}}	& \textbf{TMM\cite{hu2019progressive}}			& \textbf{Proposed}	\\
			\hline
			\hline
			\textbf{Enc.}		&\text{91x {\color{blue} (-60\%) } }		& \text{51x {\color{blue} (-29\%) }}		& \text{N/A}				& \text{36x}  \\
			\hline
			\textbf{Dec.}		&\text{230x {\color{blue} (-24\%) }}		& \text{191x {\color{blue} (-9\%) }}		& \text{207x {\color{blue} (-16\%) }}				& \text{174x}  \\
			\hline			
		\end{tabular}
	
		}
	
	\end{center}
	\label{table_complexity}
	\vspace{-7mm}
	
\end{table}

\subsection{Coding Complexity Analysis}
In addition to the coding gain, we also evaluate the coding time in Table \ref{table_complexity}. The evaluation is executed on Intel Core i7-7820X CPU@3.60GHz with 32GB memory. Same as previous works, we measure the time under the CPU platform. When using the proposed model, 36x and 174x encoding and decoding complexity is cost. Compared with \cite{li2018fully}, 60\% encoding and 24\% decoding complexity can be reduced. Compared with \cite{dumas2019context}, 29\% encoding and 9\% decoding complexity can be decreased. Compared with \cite{hu2019progressive}, 16\% decoding complexity can reduced. The relatively low complexity mainly comes from our network with few nodes or filters.

\section{Conclusions}
\label{sec:typestyle}

This paper proposes a fully NM based intra coding. First, we propose the network for all the 35 modes of variable block sizes from 4$\times$4 to 32$\times$32. Second, we propose a coding framework with NM based on the best mode probability analysis. The experimental results show that we can outperform the previous works in terms of coding gain and complexity. For the future work, we will extend the method to the next-generation video coding standard VVC.

\bibliographystyle{IEEEbib}
\bibliography{refs}

\end{document}